\documentclass{PoS}
\usepackage{aas_macros}
\title{Resolving the jet in Cygnus A}

\ShortTitle{Resolving the jet in Cygnus A}

\author{\speaker{U.~Bach}$^a$, T.~P.~Krichbaum$^a$, E.~Middelberg$^b$, W.~Alef$^a$,
and J.~A.~Zensus$^a$\\
        $^a$Max-Planck-Institut f\"ur Radioastronomie, Bonn,
Germany\\$^b$Astronomisches Institut - Ruhr
Unsiversit\"at, Bochum, Germany\\
        E-mail:
\email{ubach@mpifr-bonn.mpg.de}, \email{tkrichbaum@mpifr.de},
\email{middelberg@astro.rub.de}, \email{walef@mpifr.de}, and \email{azensus@mpifr.de}}

\abstract{Cygnus\,A is the closest ($z=0.057$) strong FR\,II radio galaxy and
therefore a key object for detailed studies of its prominent double sided jet and
nucleus. Owing to the large inclination of the jet with respect to the  observer
($>75^\circ$), and correspondingly reduced relativistic effects which allow to
measure directly the jet speed, Cyg\,A is an ideal candidate for detailed studies
of its jet physics, which is thought to be similar to those in the more luminous
quasars. Our previous studies revealed a good kinematic model for the jet of Cygnus
A, but the counter-jet speed  is still not well constrained. The central engine and
part of the counter-jet of Cyg A are likely to be  obscured by free-free absorbing
material, presumably a thick torus. At mm-wavelengths, the absorber  becomes
optically thin, which provides a more detailed view into the inner nuclear region.
Knowing the  speed of jet and counter-jet and their flux density ratio allows to
determine the jet Lorentz factors and  orientation. Therefore we started to monitor
Cyg A with global VLBI at 43\,GHz in Oct.\ 2007. Our first  epoch reveals a
previously unseen gap between both jets. This could be either a sign for a new
counter-jet component that is slowly separating or we start to see the very inner
acceleration region of  the jet which is not efficiently radiating at radio
wavelengths. Further more the image shows transversely resolved jet structures at
distances beyond $\sim 0.5$\,pc which facilitate more detailed investigations
addressing jet stratification. Analysis of the resolved jet structure shows that
the initially wide jet (opening angle $\sim 10^\circ$) collimates within the first
parsec into a edge-brightened jet with an opening angle of $\sim 3^\circ$.}

\FullConference{The 9th European VLBI Network Symposium on The role of VLBI in the
Golden Age for Radio Astronomy and EVN Users Meeting\\
		 September 23-26, 2008\\
		 Bologna, Italy}

\begin{document}

\paragraph{Observations} To obtain the highest possible sensitivity and resolution
a global VLBI array of 15 antennas, including the VLBA, GBT, VLA1, Effelsberg,
Onsala, and Noto, at 43\,GHz with a recording rate of 512\,Mbps is used to image
Cygnus A on four successive epochs separated by about 8\,month. The first epoch was
observed in October 2007, which will be followed by observations in October 2008,
February 2009, and October 2009. The data from the first epoch was correlated at
JIVE. Standard data reduction was applied in AIPS, imaging and model fitting was
done in Difmap. Except the non detection of fringes to Noto due to a fault of a
synthesizer, all stations worked well and the a priori calibration was good within
10\% to 20\% for most of the time.

\begin{figure}
\centering
\includegraphics[angle=-90,width=0.72\textwidth]{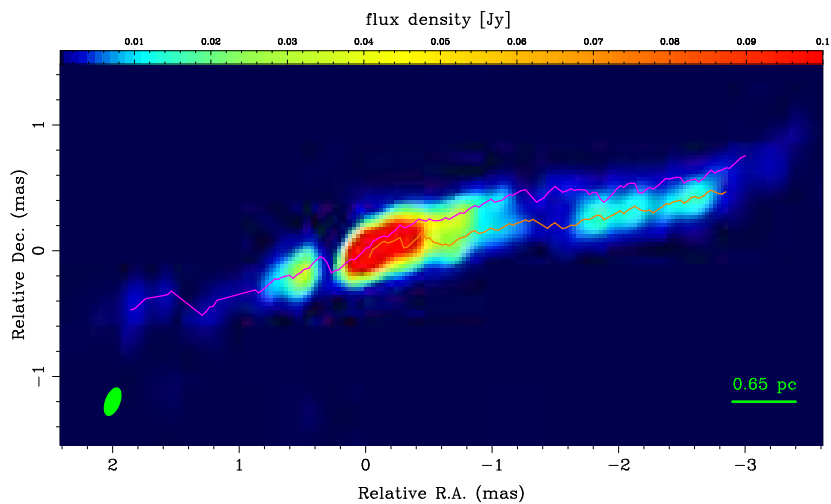}
\caption{Global 7mm VLBI image from Oct. 2007 (GB060). The image clearly shows that
the jet is transversely resolved and for some part of it, two ridge-lines along the
edges of the jet can be fitted (overlaid lines in pink and orange, see text for
details). With a peak flux density of 180\,mJy and an off-source rms noise of
0.2\,mJy the dynamic range is about 900 to 1. Noticeable is also the very
sharp emission gap at around 0.2\,mas, which separates the jet from the
counter-jet. The nature of the gap is not clear yet, but could well mark the
location of the central engine.}\label{jet} 
\end{figure}

\begin{figure}
\centering
\hbox{
\includegraphics[angle=-90,width=0.5\textwidth]{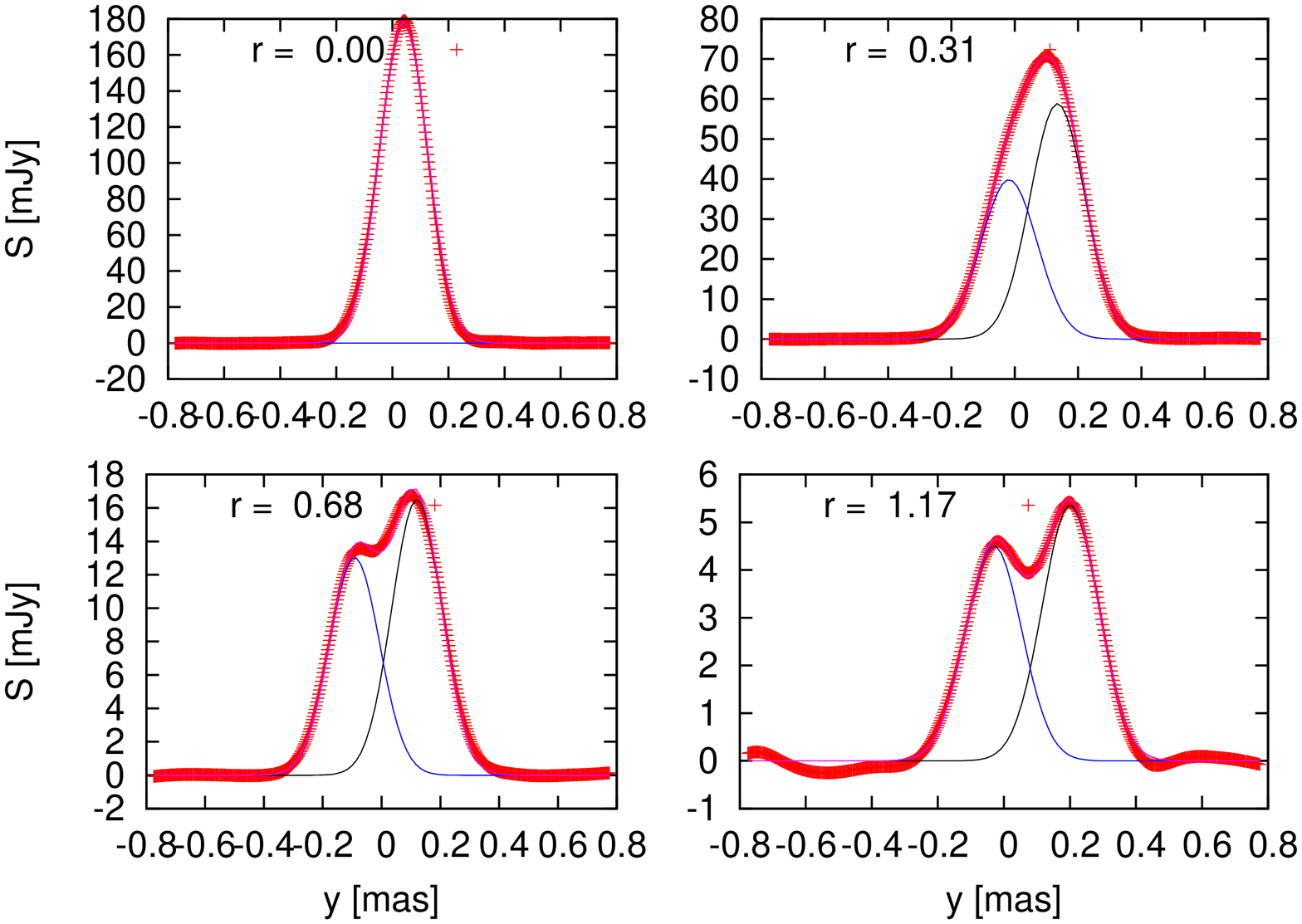}
\includegraphics[angle=-90,width=0.5\textwidth]{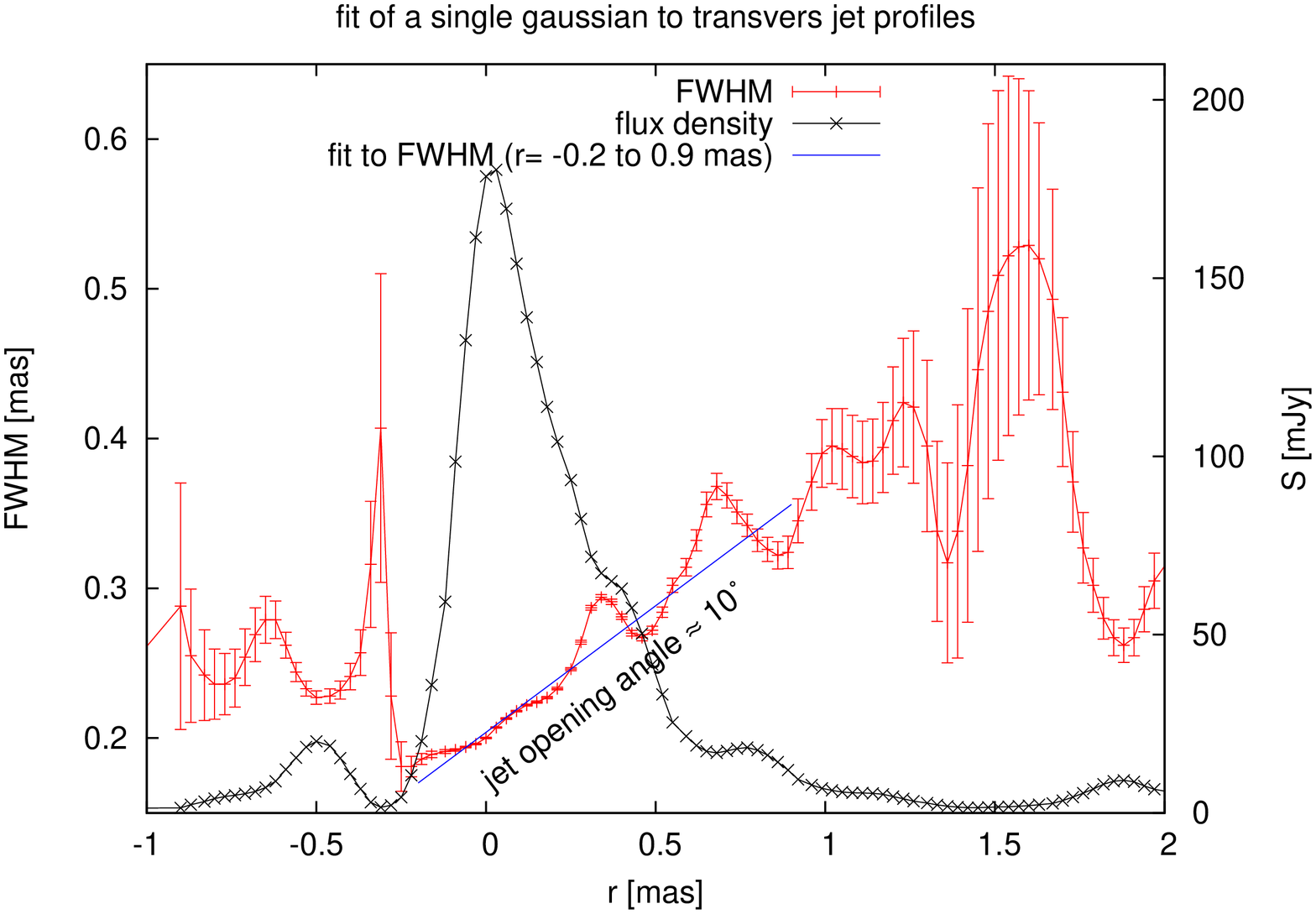}}
\caption{{\bf Left:} Flux density profiles taken transversely to the jet direction.
The jet is
clearly resolved at distances larger than 0.5 mas away from the VLBI core. To
parameterize the flux density profiles two Gaussian were fitted to the data (blue
and black line, pink is the sum of both). {\bf Right:} The jet opening angle is
obtained from the width of the Gaussian fit to the transverse profiles. For
comparison the flux density profile along the jet axis is shown
(scaleon right axis).}\label{struc}
\end{figure}

\paragraph{Results and Analysis} As seen in Fig.~\ref{jet} the jet is transversely
resolved at separations of more than 0.2 mas from the VLBI core. To measure the
width of the jet the task SLICES in AIPS was used to obtain transverse flux density
profiles of the jet and the counter-jet. Slices were taken at every pixel
(0.03\,mas) from 2.5\,mas to -3.5\,mas relative RA. Some example profiles are shown
in the left panel of Fig.~\ref{struc}. The first image (top, left panel) at
$r=0.0$\,mas shows the still compact inner jet, but with larger separations from
the core the jet broadens and shows a limb-brightened structure (top right, bottom
left to right). To parameterize the flux density profiles two Gaussian components
were fitted to the data. The core and the very inner jet are well represented by a
single Gaussian. At about 0.3\,mas to 0.5\,mas from the core the profiles are
better represented by two Gaussian and both are slowly separating from each other
with increasing distance from the core. The two ridge-lines show in Fig.~\ref{jet}
correspond to the peak positions of the Gaussian fits and follow nicely the bright
rims of the jet. For comparison all profiles were also fitted with a single
Gaussian. The flux density profile along the jet and the jet width derived from the
single Gaussian fit against core separation are shown in Fig.~\ref{struc} (right
panel). Remarkable in the flux density profile is the sharp emission gap that seems
to separates the jet from the counter-jet. It is also the region were the jet is
most compact. The nature of the gap is not clear yet. Possibilities include strong
absorption by the inner torus/accretion-disc or that it is the region where the
jets are formed and where they are less radiative. The jet width derived from the
analysis of the jet ridge-line shows a strong increase of the jet width in the
inner jet ($<0.8$\,mas) that corresponds to a jet opening angle of $\sim 10^\circ$.

The opening angle of the more extended jet can be derived from the increasing
separation of the northern and southern ridge-lines (Fig.~\ref{jet}). Here we obtain
an angle of only $2-3^\circ$. This suggests that we start to resolve the initial
jet collimation region. Interestingly the collimation appears in the inner 1.5\,pc
in agreement to the region where the jet seems to be accelerated
\cite{2005ASPC..340...30B}. A similar structure is observed also in M87
\cite{1999Natur.401..891J,2007ApJ...668L..27K}. The further analysis, specially the
linear polarization images, the following epochs, and the comparison to global
3\,mm images \cite{krichbaum08} should show if e.g. the gap
really marks the position of the central engine and if we can see here into the
region where jets are formed and accelerated
\cite{2004ApJ...605..656V,2008Natur.452..966M}.  However, without question, this
observations should facilitate more detailed investigations and allow a better
comparison to MHD simulations like to those recently done for e.g.\ M87, 3C\,273
\cite{2001Sci...294..128L}, and 0836+714 \cite{2007A&A...469L..23P}.

%\bibliographystyle{aa} % style aa.bst
%\bibliography{references}

\end{document}